\documentclass[twocolumn]{aastex62}

\usepackage{hyperref}

\newcommand{\ltsima}{$\; \buildrel < \over \sim \;$}
\newcommand{\simlt}{\lower.5ex\hbox{\ltsima}}

\newcommand{\hii}{H~{\sc ii}}
\def\arcdeg{\hbox{$^\circ$}}
\def\arcmin{\hbox{$^\prime$}}
\def\arcsec{\hbox{$^{\prime\prime}$}}

\definecolor{forestgreen(web)}{rgb}{0.13, 0.55, 0.13}

\newcommand{\aref}[1]{\hyperref[#1]{Appendix~\ref{#1}}}

\shorttitle{Stellar Feedback Evolution}
\shortauthors{Olivier et al.}

\begin{document}

\title{Evolution of Stellar Feedback in \hii\ Regions}

\author{Grace M. Olivier}
\affil{Department of Astronomy, The Ohio State University, 140 W. 18th Ave., Columbus, Ohio 43210, USA}
\affil{Center for Cosmology and AstroParticle Physics, The Ohio State University, 191 W. Woodruff Ave., Columbus, OH 43210, USA}

\author{Laura A. Lopez}
\affil{Department of Astronomy, The Ohio State University, 140 W. 18th Ave., Columbus, Ohio 43210, USA}
\affil{Center for Cosmology and AstroParticle Physics, The Ohio State University, 191 W. Woodruff Ave., Columbus, OH 43210, USA}
\affil{Niels Bohr Institute, University of Copenhagen, Blegdamsvej 17, 2100 Copenhagen, Denmark}

\author{Anna L. Rosen}
\affil{Center for Astrophysics $|$ Harvard \& Smithsonian, 60 Garden St, Cambridge, MA 02138, USA}

\author{Omnarayani Nayak}
\affil{Space Telescope Science Institute, 3700 San Martin Dr, Baltimore, MD 21218}

\author{Megan Reiter}
\affil{UK Astronomy Technology Centre, Blackford Hill, Edinburgh, EH9 3HJ, UK}

\author{Mark R. Krumholz}
\affil{Research School of Astronomy and Astrophysics, Australian National University, Canberra, ACT 2611}
\affil{ARC Centre of Excellence for Astronomy in Three Dimensions (ASTRO-3D), Canberra, ACT 2601 Australia}

\author{Alberto D. Bolatto}
\affil{Astronomy Department and Laboratory for Millimeter-wave Astronomy, University of Maryland, College Park, MD 20742, USA}

\email{olivier.15@osu.edu}

\begin{abstract}

Stellar feedback is needed to produce realistic giant molecular clouds (GMCs) and galaxies in simulations, but due to limited numerical resolution, feedback must be implemented using subgrid models. Observational work is an important means to test and anchor these models, but limited studies have assessed the relative dynamical role of multiple feedback modes, particularly at the earliest stages of expansion when \hii\ regions are still deeply embedded. In this paper, we use multiwavelength (radio, infrared, and X-ray) data to measure the pressures associated with direct radiation ($P_{\rm dir}$), dust-processed radiation ($P_{\rm IR}$), photoionization heating ($P_{\rm HII}$), and shock-heating from stellar winds ($P_{\rm X}$) in a sample of 106 young, resolved \hii\ regions with radii $\lesssim$0.5~pc to determine how stellar feedback drives their expansion. We find that the $P_{\rm IR}$ dominates in 84\% of the regions and that the median $P_{\rm dir}$ and $P_{\rm HII}$ are smaller than the median $P_{\rm IR}$ by factors of $\approx 6$ and $\approx 9$, respectively. Based on the radial dependences of the pressure terms, we show that \hii\ regions transition from $P_{\rm IR}$-dominated to $P_{\rm HII}$-dominated at radii of $\sim$3 pc. We find a median trapping factor of $f_{\rm trap} \sim$ 8 without any radial dependence for the sample, suggesting this value can be adopted in sub-grid feedback models. Moreover, we show that the total pressure is greater than the gravitational pressure in the majority of our sample, indicating that the feedback is sufficient to expel gas from the regions. 

\end{abstract} 

\keywords{Galaxy formation: Stellar feedback -- Star formation: Star forming regions --  H II regions: Compact H II region}

\section{Introduction}

Stellar feedback -- the injection of energy and momentum by stars into the interstellar medium (ISM) -- originates at the small scales of individual stars and star clusters ($\lesssim$1~pc), yet it shapes the ISM on large scales ($\gtrsim$1~kpc). Stellar feedback is responsible for the low observed star formation efficiencies in Milky Way giant molecular clouds (GMCs; e.g., \citealt{zuckerman74, krumholz07, evans09, heiderman10, murray11a, evans14, lee16, barnes17, vutisalchavakul16}) and across GMCs in nearby galaxies (e.g., \citealt{longmore13, kruijssen14, usero15, bigiel16, leroy17, gallagher18, utomo18}). This inefficiency arises from feedback processes that dissolve star clusters (e.g., see review by \citealt{krumholz19}) and ultimately destroy their host clouds (e.g., \citealt{whitworth79,matzner02,krumholz06,dale13}).

Stars have several feedback mechanisms: e.g., radiation, photoionization, stellar winds, supernovae (SNe), protostellar jets, and cosmic rays (see reviews by \citealt{krumholz19} and \citealt{rosen20a}, and references therein). Extensive recent efforts have aimed to incorporate many feedback modes in simulations of star-forming cores and GMCs (e.g., \citealt{dale14, rosen14,rosen16, dale17, tanaka17,tanaka18, rosen20b}), and in galaxy formation models (e.g. \citealt{stinson13, agertz16, hopkins18}). While SNe may be the dominant mechanism in shaping galaxies on large scales (e.g., \citealt{hopkins18}), pre-SN feedback from the other mechanisms is crucial to reproduce observed GMC properties (e.g., \citealt{grisdale18,fujimoto19}).

Observational studies are crucial to anchor and test simulations. Individual feedback modes have been assessed for large samples of sources (e.g., \citealt{rosero19}), and measurements of the relative role of multiple feedback modes have been done for particular star-forming regions 
(e.g. \citealt{lopez11,pellegrini11,ginsburg17,lee19,xu19}) and across many regions (e.g. \citealt{lopez14,chevance19,kruijssen19,mcleod19,barnes20,mcleod20}). Among these latter works, \citet[hereafter L14]{lopez14} analyzed multiwavelength data of 32 \hii\ regions in the Large (LMC) and Small Magellanic Clouds (SMC) to calculate the pressures associated with direct radiation, dust-processed radiation, photoionization heating, and shock-heating from stellar winds and SNe. They found that the warm ($10^{4}$~K) gas pressure is the dominant feedback mechanism at the \hii\ region shells. \cite{mcleod19} also examined two LMC star-forming complexes using integral-field data from MUSE, characterizing the stellar content, the gas properties, and the kinematics. Consistent with the L14 results, \cite{mcleod19} determined that photoionization heating drives the dynamics in their sample. 

The studies published to date have focused on relatively large ($R\gtrsim \mbox{few}$ pc) and evolved ($t\gtrsim \mbox{few}$ Myr) \hii\ regions (except the recent work by \citealt{barnes20}). However, theoretical models \citep[e.g.,][]{krumholz09,geen20} suggest that mechanisms other than photoionization may be comparatively more important early in the evolution of \hii\ regions, when they are significantly smaller and younger than the range probed by L14 and other works. To assess this possibility, in this paper we aim to explore how feedback properties differ at earlier stages in massive star formation and how the driving feedback mechanisms evolve with time. We therefore carry out an analysis similar to that of L14 but targeting a radio/IR-selected sample of much smaller, younger \hii\ regions.

In~\autoref{sec:data}, we describe our sample and data used in our analysis.  In~\autoref{sec:methods}, we review our methods for measuring each form of feedback. In~\autoref{sec:results}, we present our results, and in~\autoref{sec:discuss}, we discuss the implications regarding young \hii\ region dynamics, their evolution, and how our results can inform stellar feedback modeling in numerical simulations. We summarize our conclusions in~\autoref{sec:conclusions}.

\section{Sample and Data} \label{sec:data}

\begin{figure}
\includegraphics[width=\columnwidth]{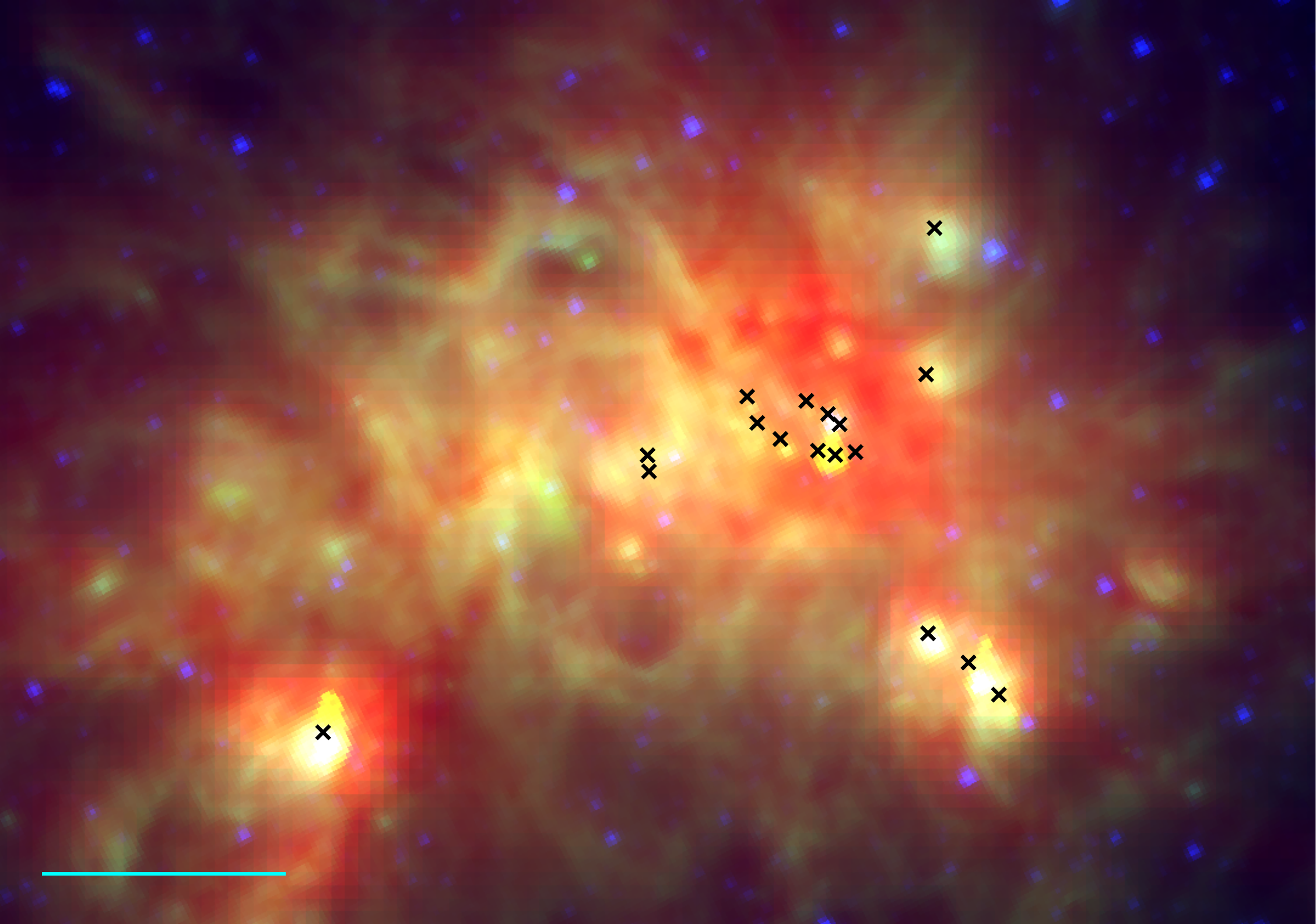}
\caption{Three-color image of the massive star-forming region W49A, with {\it Herschel} 70$\mu$m in red, GLIMPSE 8.0$\mu$m in green, and GLIMPSE 3.6$\mu$m in blue. Black Xs mark the locations of nearly twenty young \hii\ regions in our sample. The scale is 1\arcmin\ in length, $\approx$3.5~pc for a distance of 11.9~kpc to W49A \citep{quireza06}.}
\label{fig:w49a}
\end{figure}

To evaluate the role of feedback mechanisms at the earliest stages of massive star formation, we consider 128 young \hii\ regions: 49 UC\hii\ regions (defined as those with radii $R \lesssim 0.05$~pc), 67 compact \hii\ regions (with radii $0.05\lesssim R \lesssim 0.25$~pc), and 12 small \hii\ regions (with radii $0.25\lesssim R \lesssim 0.5$~pc) from the ATLASGAL survey \citep{urquhart13}. We require that the regions have bolometric luminosities $L_{\rm bol}$ measured by \cite{mottram11} and reported in \cite{urquhart13}, limiting the \cite{urquhart13} sample of 213 sources to our sample of 128 compact \hii\ regions. \cite{urquhart13} used surveys that were complete down to B0 stars on the other side of the galaxy (ATLASGAL and CORNISH; \citealt{purcell13}). The \cite{mottram11} MSX survey is complete down to \hii\ regions with $L_{\rm bol} \sim 10^4 L_\odot$ which matches the ATLASGAL survey well. Therefore, this requirement of $L_{\rm bol}$ in our study gives a representative sample of small \hii\ regions in the inner Milky Way. Out of the 128 sources, 106 were resolved in the CORNISH survey \citep{purcell13}, so we measure the feedback pressures in this sample and set limits on these terms for the 22 unresolved sources. The ATLASGAL survey covered a portion of the Milky Way Galactic plane, from longitude $10^\circ \leq l \leq 60^\circ$ and latitude $-1^\circ \leq b \leq 1^\circ$. These \hii\ regions were detected in the 5-GHz band by the CORNISH survey \citep{purcell13} and in the 870 $\mu$m band by the ATLASGAL survey and then confirmed as \hii\ regions using mid-infrared colors from the GLIMPSE survey \citep{benjamin03, churchwell09}. The \hii\ regions have well-defined radii from the 5-GHz measurements and kinematic distances derived by \cite{urquhart13}. To illustrate the sample, in \autoref{fig:w49a}, we show a three-color image of the massive star-forming region W49A, which contains nearly twenty of the regions considered in this work. 

To evaluate the pressures associated with the different feedback modes, we employ radio, infrared (IR), and X-ray observations, as detailed in \autoref{sec:methods}. Specifically, to measure the warm ($\sim$10$^{4}$~K) gas pressure ($P_{\rm HII}$) associated with photoionization, we use the 5-GHz detections from the CORNISH survey \citep{purcell13}. To constrain the dust-processed radiation pressure ($P_{\rm IR}$), we utilize 2MASS J, H, and, K-band photometry \citep{skrutskie06}, 8, 12, 14, and 21.3-$\mu$m data from the RMS catalog \citep{lumsden13}, 70-$\mu$m data from {\it Spitzer} MIPS \citep{carey09}, 60- and 100-$\mu$m data from IRAS \citep{iras88}, and 870 $\mu$m data from ATLASGAL \citep{urquhart13}. We require that our sample has 21.3-$\mu$m data in order to constrain the peak of the spectral energy distribution (SED). To assess the hot ($\sim$10$^{7}$~K) gas pressure attributed to shock-heating by stellar winds ($P_{\rm X}$), we analyze 0.5--7~keV archival data from the {\it Chandra} X-ray Observatory. 

\section{Methods} \label{sec:methods}

To measure the pressures associated with each feedback mode, we adopt methods similar to those of \cite{lopez11} and L14, with some differences described in the following sections. As the sources in our sample are not highly resolved, we measure integrated pressures averaged within the \hii\ region shells. We adopt \hii\ region radii $R$ from \cite{urquhart13} using distances measured with galactic kinematics and maser parallaxes. The regions have $R = 0.01-0.4$~pc (angular radii of $R_{\rm ang} = 0.5-11.7$\arcsec) and are at distances of $D=1.1-18.8$ kpc. 

\subsection{Direct Radiation Pressure}

Following L14, we define the direct radiation pressure as the momentum available to drive motion in the \hii\ region shells at a radius $R$ from the central stars:

\begin{equation}
    P_{\mathrm{dir}} = \frac{3 L_{\mathrm{bol}}}{4\pi R^2 c},
    \label{eq:pdir}
\end{equation}

\noindent
where $L_{\rm bol}$ is the bolometric luminosity of the central stars\footnote{Note that the factor of 3 in the numerator of \autoref{eq:pdir} arises because we are computing the volume-averaged radiation pressure within the \hii\ region rather than simply the pressure at the surface. The reason for computing the volume-averaged pressure is that $\int P_{\rm dir} \, dV$ is the quantity that appears in the virial theorem describing the overall dynamics of the region \citep{McKee92a}. For more discussion of why it is important to include this factor, see L14.}. \cite{mottram11} derived $L_{\rm bol}$ for a subsample of the \hii\ regions by fitting SED models of young stellar objects (YSOs) to their SEDs \citep{robitaille07}. As noted by \cite{urquhart13}, the ionizing fluxes they measured are consistent with these $L_{\rm bol}$ from \cite{mottram11}. Thus, we adopt the $L_{\rm bol}$ reported by \cite{mottram11} for our \hii\ regions.  

\subsection{Dust-processed Radiation Pressure}

The stellar radiation is absorbed by dust and thermally re-radiated at longer wavelengths in the IR, thereby enhancing radiation pressure in young star-forming regions. Measuring the dust-processed radiation pressure, $P_{\rm IR}$, for our sources is significantly more challenging than for the larger \hii\ regions examined in L14 and similar studies. For evolved sources, the column of material inside and in front of the \hii\ region is small enough to be optically thin at far-IR and longer wavelengths, and thus one can use dust re-radiation at these wavelengths as a diagnostic of the radiation field seen by the grains. We have intentionally selected much more embedded sources, which may be optically thick in the far-IR regime. Consequently, we estimate the volume-averaged radiation pressure by modelling the complete near-IR to sub-mm SED and then deriving the pressure from the model. 

For this purpose, we fit the IR data using synthetic SEDs from models of young stellar objects (YSOs) computed by \cite{robitaille17} to find a geometry that produces the IR SED from each region. In an effort to recreate the models adopted by \cite{mottram11} from \cite{robitaille07}, we use the most similar model set out of the 18 available in \cite{robitaille17}, the \textit{spubsmi} model set. All model sets include a central star, an option of a passive (non-accreting) disk, a power-law or an accreting, rotationally flattened envelope \citep{ulrich76}, an internal bipolar cavity due to outflows, an ambient medium, and the option of the inner radius of the disk and envelope as either the dust sublimation radius or as a variable. Each model set has 14 parameters that are considered in the fit: stellar radius, stellar temperature, disk mass, disk inner radius, disk outer radius, the power-law index describing disk flaring, disk surface density power, disk scale height, envelope density, the power-law index describing the envelope density profile, envelope centrifugal radius, cavity density, cavity opening angle, and the power-law exponent of the cavity opening.

We chose the \textit{spubsmi} model set because high-mass YSOs have observed accretion disks \citep{patel05,kraus10}; however, we note that no \cite{robitaille17} model set has accreting disks, arguing that the difference between a passive and an accreting disk would not noticeably change the SED. We tested different model sets on a sample of our regions, and the \textit{spubsmi} set fit both the near-IR and far-IR data the best in terms of the resulting $\chi^2$ values.

We fit the synthetic SEDs produced by this model set to the IR data using the SEDfitter from \cite{robitaille07}.  We used the same extinction law as \cite{forbrich10} which has $R_{\rm V}$ = 5.5 based on the larger grains anticipated in dense star-forming regions. When fitting the data, we limited the distances to within 2 kpc of the values reported by \cite{urquhart13}. The distance functions mostly as a normalization of the models during fitting: by limiting the distances, we exclude models that have incident fluxes indicative of regions either much farther or much closer than are observed. In addition to this normalization aspect of distance, more distant regions have more material in the beam, so fitting with a single star model would be a worse assumption. Nevertheless, the observed flux is dominated by the brightest source, and with our adequate fits, the assumption seems reasonable. None of the regions have clearly resolved, multiple stars, and from the SED fits, it is not possible for us to distinguish whether the regions are powered by one or more stars.

\begin{figure}
    \includegraphics[width=\columnwidth]{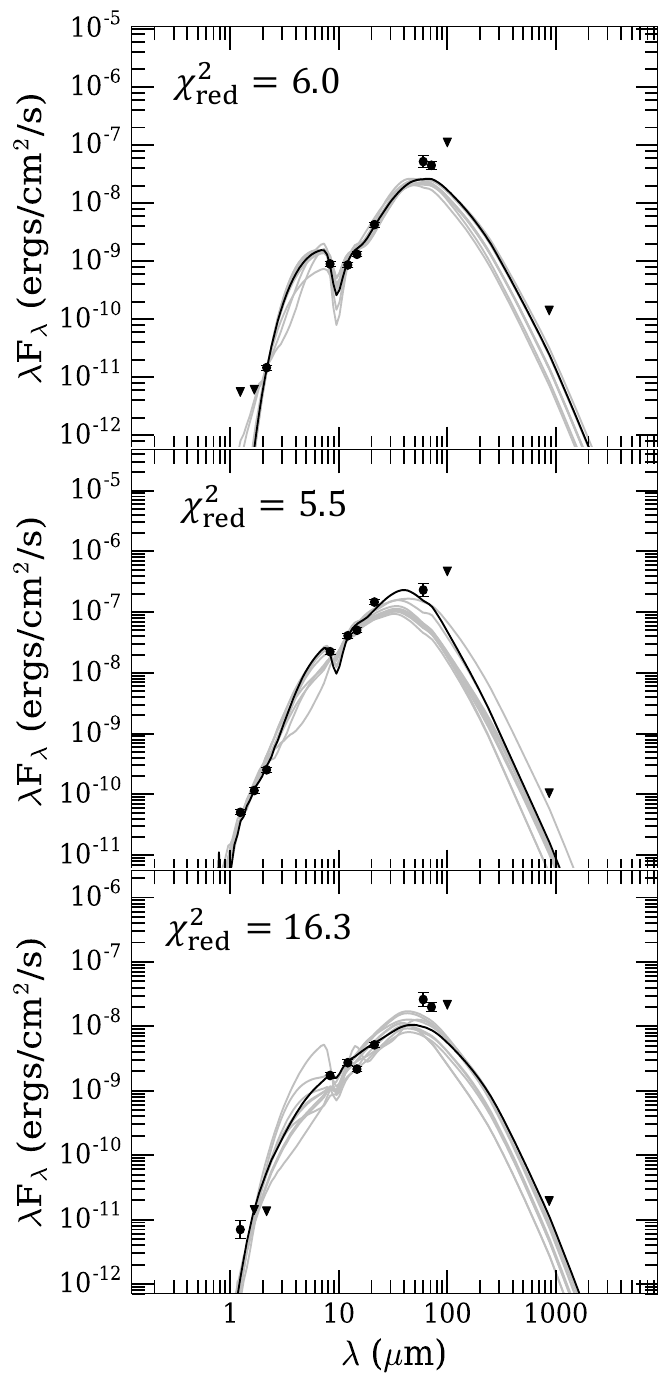}
    \caption{Examples of measured SEDs (circles with error bars represent detections, downward triangles are upper limits), together with the SEDfitter output of the ten best-fit models for three sample \hii regions.  The black lines are the best-fit models, and the gray lines are the nine next best-fit models. The best-fit $\chi^2$ of the three regions is reported in the plot.}
    \label{fig:SED}
\end{figure}

We fit the \cite{robitaille17} models to an SED consisting of measurements at J, H, and K-band (from 2MASS), 8, 12, 14, and 21 $\mu$m (from MSX), 60 and 100 $\mu$m (from IRAS), 70 $\mu$m (from \textit{Spitzer}/MIPS), and 870 $\mu$m from ATLASGAL, as described in \autoref{sec:data}. For the J, H, K, and 70-$\mu$m measurements, we compare to the filter-convolved fluxes in the \citet{robitaille17} models, while for the other wavelengths, we use the flux at the central wavelength of the filters. Additionally, the apertures of the measurements are necessary for SEDfitter to provide an accurate fit We treat the measurements at 100 and 870 $\mu$m as upper limits, since the beam sizes for these measurements are much larger than the $\approx 5$\arcsec~sizes of our sample \hii\ regions. We set the apertures for the data shorter than 21 $\mu$m, at 60 $\mu$m, at 70 and 100 $\mu$m, and at 870 $\mu$m to 3\arcsec, 10\arcsec, 20\arcsec, and 55\arcsec, respectively.  This set of measurements ensures good coverage on both sides of the peak of the SED, enabling strong constraints on the model fits. We show some example fits produced by this procedure in \autoref{fig:SED}.

In our analysis, we include only the regions where the reduced chi-squared from the fit is $\chi^2_{\rm red}<30$ which limits our sample to 81 sources. \cite{nayak16} used a $\chi^2_{\rm red}$ cut of 10 to ensure that only YSOs were included in their sample.  We use a less stringent cut on $\chi^2_{\rm red}$ because \cite{urquhart13} already confirmed these objects are \hii\ regions.

Once we determined the best-fit SED model, we took the 14 parameters and recreated the geometry in HYPERION, a 3D Monte-Carlo dust continuum radiative transfer code, to obtain the dust temperature distribution \citep{robitaille11}. HYPERION outputs an axi-symmetric dust temperature $T_{\rm d}$ map as heated by the radiation from the central source. From the $T_{\rm d}$ distribution, we calculate $P_{\rm IR}$ for each source by integrating over the two-dimensional axi-symmetric grid of $T_{\rm d}$:

\begin{equation}
    P_{\mathrm{IR}} = \frac{1}{V} \int_0^R \int_0^\pi \frac{1}{3} a T_{\rm d}^4 2 \pi r^2 \sin(\theta) d\theta dr
\end{equation}

\noindent
where $a$ is the radiation constant $a=7.56 \times 10^{-15}$ erg cm$^{-3}$ K$^{-4}$ and $V$ is the \hii\ region volume. Note that this expression implicitly assumes that the IR radiation temperature is equal to the dust temperature; this is a reasonable approximation in the highly-opaque regions with which we are concerned. We integrate out to $R$, the observed \hii\ region radius, to include the dusty shell surrounding the region. We set the dust temperature within the dust sublimation radius to $T_{\rm d} = T_{\rm sub} = 1600$ K, the dust sublimation temperature, to account for the energy contained in that radius. 

\subsection{Warm Ionized Gas Pressure} \label{subsec:warm}

The photoionization heating from massive stars creates warm, $\approx$10$^4$ K gas, the classical driver of \hii\ regions.  We measure the pressure from this warm gas by using the ideal gas law:

\begin{equation}
    P_{\mathrm{HII}} = n_{\rm HII}kT_{\rm HII}.
\end{equation}

\noindent
We adopt a warm-gas temperature of $T_{\rm HII} = 10^4$~K, and we estimate the density, $n_{\rm HII}$, to calculate $P_{\rm HII}$. $n_{\rm HII}$ is the number density of free particles and depends on the ionization state of the gas; if hydrogen is fully ionized and helium is singly ionized, then $ n_{\rm HII} = n_{\rm e} + n_{\rm H} + n_{\rm He} \approx 2 n_{\rm e}$. We use Equation~5.14b from \cite{rybicki79} for free-free emission to calculate $n_{\rm e}$:

\begin{equation}
    n_{\rm e} = \left( \frac{1.47 \times 10^{37} 4 \pi D^2 F_\nu T_{\rm HII}^{1/2}}{ {\overline g_{{\mathrm{ff}}} R^3}} \right)^{1/2}
\end{equation}

\noindent 
where $D$ is the distance to the region,  $F_{\nu}$ is the flux density at frequency $\nu$, and $V$ is the volume of the region. In this study, we use the measurements from CORNISH at 5 GHz, where free-free dominates.  We adopt a Gaunt factor of 5.1 based on \cite{draine11book} and \cite{vanhoof14}, which indicate that for a 10$^4$K gas observed at 5 GHz, the Gaunt factor is larger than the commonly adopted 1. This higher Gaunt factor lowers our $P_{\rm HII}$ measurements by a factor of 5$^{1/2}$ compared to a Gaunt factor of 1. If instead a $\overline g_{\rm ff} = 1$ was adopted, then the primary result that would change is the transition radius where $P_{\rm HII} = P_{\rm IR}$ would become $\sim$ 0.9 pc instead of $\sim$ 3 pc in \autoref{subsec:evol}.

\subsection{Hot Gas Pressure}

The stars powering the young \hii\ regions have radiatively-driven stellar winds that shock-heat gas to $\gtrsim$10$^6$ K \citep{rosen14}.  This hot gas creates pressure within the region according to the ideal gas law:

\begin{equation}
    P_\mathrm{X} = 1.9 n_{\rm X} k T_{\rm X}
\end{equation}
\noindent
where $n_\mathrm{X}$ is the electron density in the hot gas, and the factor of 1.9 assumes that He is doubly ionized and the He mass fraction is 0.25. 

Based on searches of the \textit{Chandra} Data Archive, 26 out of 128 have archival observations. Among the 26 sources, only 6 have detections with $>$10 net counts in the 0.5--7.0~keV band. In these cases, the regions are not resolved\footnote{The on-axis PSF of {\it Chandra} ACIS is 0.492\arcsec\, and the PSF 5\arcmin\ off-axis increases to $>$1\arcsec: \href{https://cxc.harvard.edu/proposer/POG/html/chap4.html}{https://cxc.harvard.edu/proposer/POG/html/chap4.html}}, and the limited number of counts precludes reliable spectral modeling. However, the high median energies of the counts ($\approx$3--6~keV) and the elevated hardness ratios\footnote{Defined as HR = [$F$(2--8 keV)$-F$(0.5--2 keV)]/[$F$(2--8 keV)$+F$(0.5--2 keV)], where $F$ is the net counts observed in a given bandpass.} ($\approx$0.5--1.0) of the targets suggest that the spectra are quite hard. Previous X-ray studies of young \hii\ regions have also found that hard X-rays dominate their emission (e.g., \citealt{tsujimoto06,anderson11}), which may indicate very hot plasma temperatures (with $kT_{\rm X} \gtrsim 6$~keV) and/or extreme column densities (of $N_{\rm H} \gtrsim10^{23}$~cm$^{-2}$) that attenuate the soft X-rays.

In young \hii\ regions, the origin of the X-ray emission is uncertain: it may be from unresolved point sources, interacting/colliding stellar winds, or wind-blown bubbles (see \citealt{tsujimoto06}). However, the available data have insufficient counts to distinguish between these scenarios. Thus, given that the X-rays may arise from these other channels and not just the shock-heated gas from individual stellar winds, we treat all of the hot-gas measurements as upper limits.

We calculate the upper-limits on $n_{\rm X}$ for each region by simulating an optically-thin thermal plasma with metallicity of $Z = Z_{\sun}$ and temperature $kT_{\rm X}$=0.4~keV. We assume a Galactic column density $N_{\rm H}$ toward each region using the values from \cite{nh}. We use WebPIMMS\footnote{\href{https://heasarc.gsfc.nasa.gov/cgi-bin/Tools/w3pimms/w3pimms.pl}{https://heasarc.gsfc.nasa.gov/cgi-bin/Tools/w3pimms/w3pimms.pl}} to obtain the normalization $norm$ of the resulting X-ray spectra, which is defined as $norm = (10^{-14}/4 \pi D^2) \: EM_{\rm X}$, where $EM_{\rm X} = \int n_{\rm X}^2 dV$. Given the $norm$ values and assuming the distance to the regions, we measure $n_{\rm X}$ by integrating over the volume of the region.

\subsection{Errors and Uncertainties in Pressure Terms} \label{sec:uncertainties}

To assess uncertainty in our pressure measurements, we examine the largest source of error in each calculation. For $P_{\rm HII}$, $P_{\rm dir}$, and $P_{\rm IR}$, the greatest uncertainty arises from the $\sim$10\% error in the \hii\ region radii reported by \citep{urquhart13}. Thus, to ascertain the confidence range in the pressure terms, we adopt radii of $R \pm0.1R$ and compute the associated pressures. For the 22 unresolved regions, we set 0.75\arcsec\ as the sources' angular diameter. These upper limits yield lower limits on the pressures derived for these objects. 

\section{Results} \label{sec:results}

From our 128 region sample, 106 sources were resolved in the CORNISH survey \citep{purcell13}, and 81 meet our requirements for the IR SED fits. We find that 68 of the 81 regions are $P_{\rm IR}$-dominated, and 3 regions are $P_{\rm dir}$-dominated. Additionally, 4 regions have $P_{\rm HII} \sim P_{\rm IR}$ (within the uncertainties), and 6 others have $P_{\rm dir} \sim P_{\rm IR}$ (within the uncertainties). Overall, the median $P_{\rm dir}$ ($P_{\rm HII}$) is 17\% (11\%) of the median $P_{\rm IR}$ in our sample.

\begin{figure*}
\includegraphics[width=\textwidth]{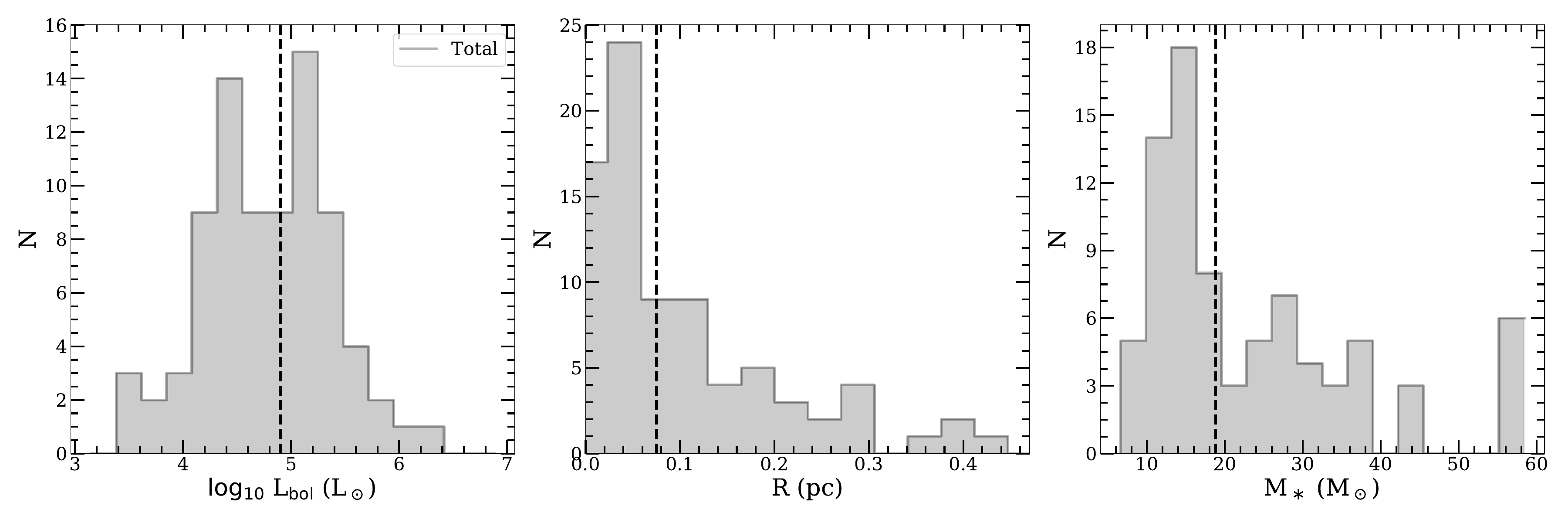}
\caption{Histograms of properties of the young \hii\ regions. The vertical dashed lines represent the medians of the parameters. $L_{\rm bol}$ of the sample was measured by \cite{mottram11}, $R$ was reported in the ATLASGAL survey \citep{urquhart13}, and we derive $M_{\ast}$ from $L_{\rm bol}$ (see \autoref{sec:results}).} 
\label{fig:histograms}
\end{figure*}

We plot the distribution of $L_{\rm bol}$, radius $R$, and central star mass $M_{\ast}$ for all of the regions in \autoref{fig:histograms}. We derive $M_{\ast}$ using $L_{\rm bol}$ and the empirical measurements of O stars from \cite{martins05} and of B stars from \cite{schmidtkaler82}. We report the parameter ranges covered by the four groups of \hii\ regions (with $P_{\rm HII} \sim P_{\rm IR}$, $P_{\rm dir} \sim P_{\rm IR}$, $P_{\rm dir}$-dominated, and $P_{\rm IR}$-dominated regions) in \autoref{tbl1}. For the 81 region sample, the median $\log L_{\rm bol} / L_{\odot}$ is 4.9, the median $R$ is 0.08 pc, and the median $M_{\ast}$ is 18.8 $M_{\odot}$.

\begin{figure}
\includegraphics[width=\columnwidth]{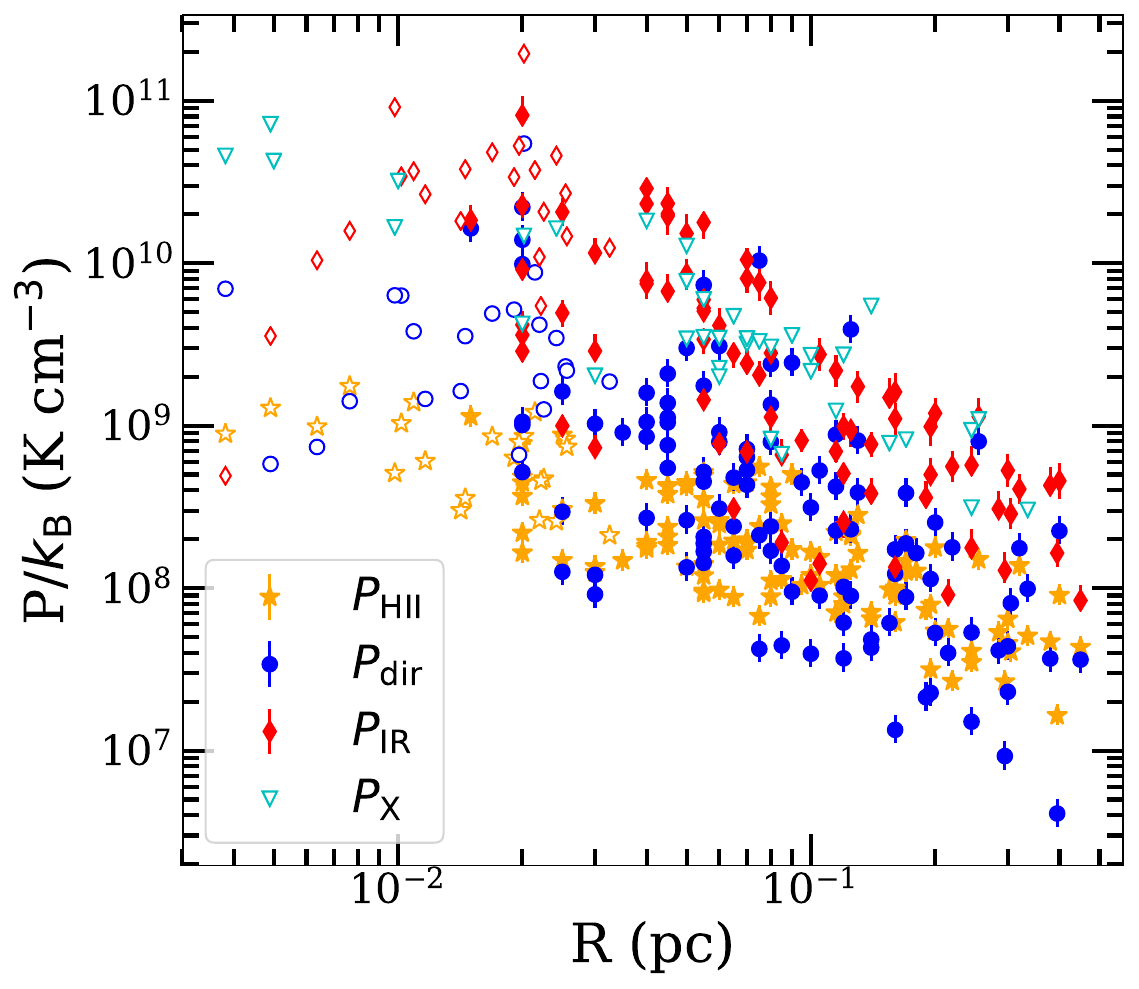}
\caption{The young \hii\ region pressure terms versus radius $R$. The open stars and circles represent lower-limits on $P_{\rm HII}$ and $P_{\rm dir}$, respectively. The open diamonds represent upper-limits on $P_{\rm IR}$ because the regions are not resolved (see \autoref{sec:uncertainties}). The open triangles represent upper limits on $P_{\rm X}$ from the lack of detections or from uncertainties in whether the X-ray emission arises from diffuse gas or from the central star surface.}
\label{fig:pvr}
\end{figure}

\begin{figure}
\includegraphics[width=\columnwidth]{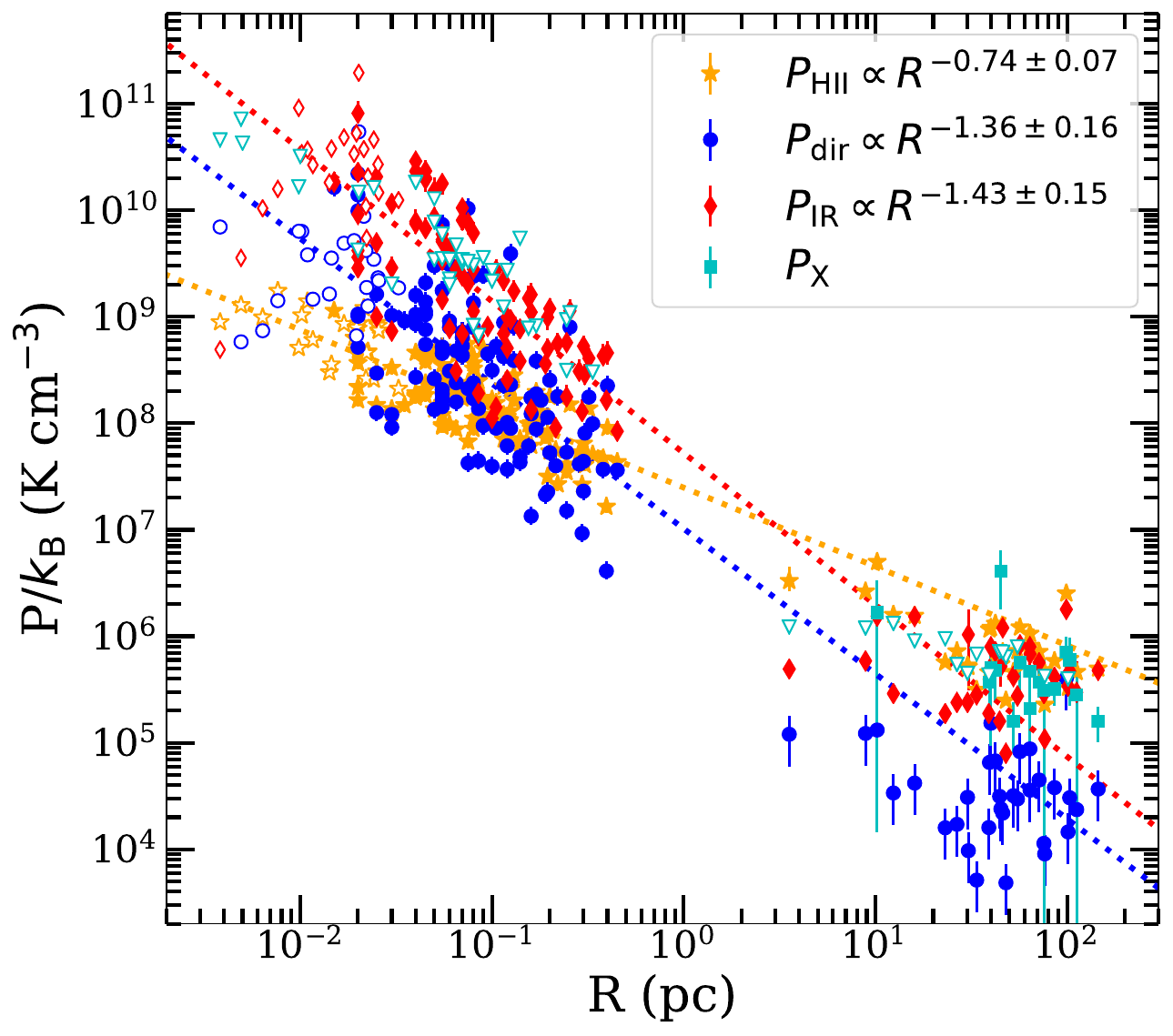}
\caption{Feedback pressures versus radius $R$ for the young \hii\ region sample (upper left) and the evolved \hii\ region sample from L14 (bottom right). The open stars and circles represent lower limits in $P_{\rm HII}$ and $P_{\rm dir}$, and the open diamonds represent upper limits on $P_{\rm IR}$ due to unresolved regions (see \autoref{sec:uncertainties}). The open triangles represent upper limits on $P_{\rm X}$. The dotted lines show the best-fit radial trends in the pressure terms of the young \hii\ region sample.}
\label{fig:pvr_wLopez14}
\end{figure}

In \autoref{fig:pvr}, we plot the pressures derived for the sample, and in ~\autoref{fig:pvr_wLopez14}, we compare our results with those of the evolved \hii\ regions in the LMC and SMC from L14\footnote{We have recomputed the $P_{\rm HII}$ measured by L14, changing the Gaunt factor of ${\overline g}_{{\mathrm{ff}}}=1.2$ in that work to ${\overline g}_{{\mathrm{ff}}}=4.8$.}. The pressure terms decrease with radius, as expected given the increasing volume as the shell expands. L14 found that $P_{\rm HII}$ dominated for the evolved \hii\ regions, while $P_{\rm IR}$ is nearly an order of magnitude lower in their sample, contrasting the results from the young \hii\ regions in our sample.

\begin{deluxetable*}{rcllll}
\tablecaption{Properties of the 4 Groups of \hii\ regions Sorted by the Dominant Pressure Terms}
\tablehead{\colhead{Property} & \colhead{Units} & \colhead{$P_\mathrm{HII} \sim P_\mathrm{IR}$}  & \colhead{$P_\mathrm{dir} \sim P_\mathrm{IR}$} & \colhead{$P_\mathrm{dir}$}  & \colhead{$P_\mathrm{IR}$}  }
\startdata	
Number                  &                       & 4          & 6            & 3        & 68       \\ 
\hline
log $L_\mathrm{bol}$    & L$_\odot$             & 4.1--4.9   & 5.0--6.4     & 5.3--5.8     & 3.5--6.2       \\
$M_\ast$               	& M$_\odot$			    & 13.0--18.8 & 20.8--58.0   & 31.0-58.0    & 9.8--58.0      \\
$f_\mathrm{trap}$		&       			    & 2.6--5.3   & 1.7--2.6     & 1.2--1.8     & 2.8--121       \\
$R$		                & pc				    & 0.09--0.22 & 0.02--0.26   & 0.02--0.16   & 0.02--0.45  	\\
$r_\mathrm{ch}$     	& pc            	    & 0.01--0.04 & 0.03--0.46   & 0.03--0.17   & 0.01--13.5  	\\
$R$/$r_\mathrm{ch}$ 	&	                    & 3.3--11.66 & 0.24--2.11   & 0.47--2.89   & 0.01--5.51 	\\
\enddata 
\tablecomments{Range of values from the sample of 81 young \hii\ regions that have measurements for $P_\mathrm{HII}$, $P_\mathrm{dir}$, and $P_\mathrm{IR}$. \looseness=-2}
\label{tbl1}
\end{deluxetable*}

Our results, in combination with those found in L14, can probe the transition from radiation pressure dominated to warm gas pressure dominated \hii\ regions. Direct radiation pressure, $P_{\rm dir}$, and $P_{\rm HII}$ have different radial dependencies in a single region: $P_{\rm dir} \propto r^{-2}$, whereas $P_{\rm HII} ~ r^{-3/2}$. The radial dependence of $P_{\rm IR}$ is more complex, since it depends on the opacity of the dusty material around the \hii\ region as well as its radius, but is also in general expected to fall sharply with $R$. Consequently, a region transitions from being radiation-pressure driven to gas-pressure driven at a characteristic radius $r_{\rm ch}$ \citep{krumholz09}:

\begin{equation}
    r_\mathrm{ch} = 0.018 f_\mathrm{trap}^2 S_{49} \;\rm pc,
\end{equation}

\noindent
where we assume a temperature of $T_{\rm HII} = 10^{4}$~K, and $S_{49}$ is the ionizing photon rate in units of $10^{49}$ photons~s$^{-1}$.
The factor $f_{\rm trap}$ represents the amount that radiation pressure is enhanced by trapping energy within the shell from stellar winds ($f_{\rm trap,w}$), infrared photons ($f_{\rm trap,IR}$), or Ly$\alpha$ photons ($f_{\rm trap,Ly\alpha}$): 

\begin{equation}
    f_{\rm trap} = 1 + f_{\rm trap,w}+f_{\rm trap,IR}+f_{\rm trap,Ly\alpha}, 
\end{equation}

\noindent
where the 1 represents absorption of the direct radiation. We are unable to constrain $f_{\rm trap,w}$ without X-ray detections of the diffuse gas in the sources, and $f_{\rm trap,Ly\alpha} \approx 0$ since Ly$\alpha$ trapping is limited by the presence of dust \citep{henney98}. Therefore, we assume that $f_{\rm trap} \approx 1 + f_{\rm trap,IR} \equiv 1 + P_{\rm IR}/P_{\rm dir}$. 

\begin{figure*}
\includegraphics[width=\textwidth]{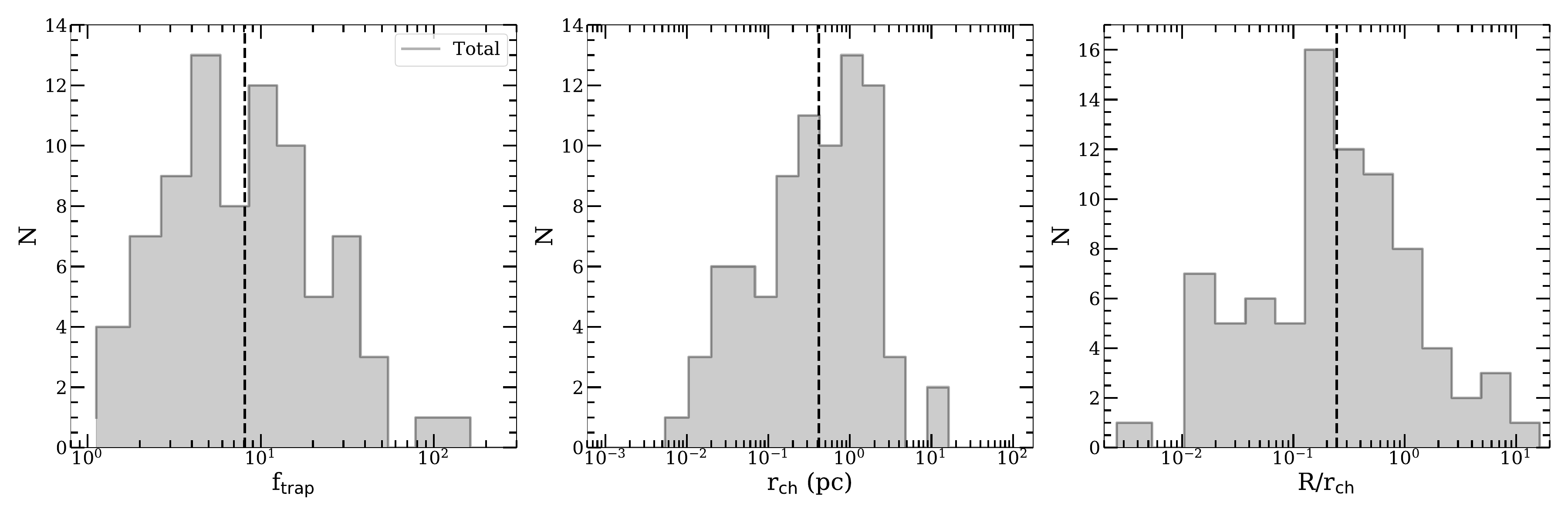}
\caption{Histograms of $f_{\rm trap}$ (the amount that radiation pressure is enhanced by the trapping energy within the shell), $r_{\rm ch}$ (the characteristic radius where a region transitions from radiation-pressure driven to gas-pressure driven), and $R/r_{\rm ch}$ (the ratio of the \hii\ region radius and the characteristic radius) for the young \hii\ regions. The vertical dashed lines represent the medians of the parameters.} 
\label{fig:histograms_ftrap}
\end{figure*}

The distributions of $f_{\rm trap}$, $r_{\rm ch}$, and $R/r_{\rm ch}$ are shown in \autoref{fig:histograms_ftrap}. The median values of these parameters are $f_{\rm trap}$ = 8.1, $r_{\rm ch}$ = 0.42~pc, and $R/r_{\rm ch}$ = 0.24.

\section{Discussion} \label{sec:discuss}

\subsection{Feedback Evolution} \label{subsec:evol} 

As shown in \autoref{fig:pvr_wLopez14}, $P_{\rm HII}$ and $P_{\rm dir}$ are of similar magnitude for the young \hii\ regions, but $P_{\rm IR}$ dominates statistically significantly in 68 out of the 81 source sample.  By contrast, in the evolved L14 sources, $P_{\rm IR}$ is roughly an order of magnitude lower than $P_{\rm HII}$, and almost all are $P_{\rm HII}$-dominated.

Moreover, the young \hii\ region pressures are orders of magnitude larger than those measured for the evolved sample from L14. This result suggests radial dependence of the pressures. For an individual \hii\ region, $P_{\rm HII} \propto R^{-3/2}$ (Equation~4), while $P_{\rm dir} \propto R^{-2}$ (Equation~1). However, we note the heterogeneity of our sample and the L14 sample. For example, based on their luminosities (see \autoref{tbl1}), the young \hii\ regions considered here are powered by individual O- and B-stars or small star clusters, whereas the L14 sample is driven by star clusters of mass $M \approx 300$--3$\times10^{4}~M_{\sun}$. Furthermore, the L14 sample is comprised of \hii\ regions in the LMC and SMC, where the metallicity is lower (with $Z \approx 0.5Z_{\sun}$ and 0.2$Z_{\sun}$, respectively; \citealt{russell92,kurt98}).

To assess the radial dependence, we measure the power-law slopes with respect to radius for each pressure term for only the young \hii\ sample. We use all 106 resolved regions to measure the slopes for $P_{\rm HII}$ and $P_{\rm dir}$, but we use only the 81 regions with $P_{\rm IR}$ measurements to fit the $P_{\rm IR}$ slope.  We find $P_{\rm dir} \propto R^{-1.36\pm0.16}$, $P_{\rm IR} \propto R^{-1.43\pm0.15}$, and $P_{\rm HII} \propto R^{-0.74\pm0.07}$. We note that $P_{\rm IR}$ was measured differently in this work compared to L14: L14 used the \cite{draine07} dust models, which did not apply to the embedded \hii\ regions analyzed here because they are optically thick.  Consequently, the best-fit power-law does not reflect the $P_{\rm IR}$ measurements of the evolved \hii\ regions.

Given the different radial trends between $P_{\rm IR}$ and $P_{\rm HII}$, we estimate the radius where the sources transition from $P_{\rm IR}$-dominated to $P_{\rm HII}$-dominated: $\sim$3~pc.  This assumes that the dust remains optically thick around the \hii\ regions out to these large radii, which may not be physically probable. If instead they become optically thin, then this transition would occur at smaller radii. In the future, observations of $\sim0.5-3$~pc radius \hii\ regions are necessary to evaluate this transition.

$P_{\rm HII}$ was the dominant form of feedback in the evolved \hii\ region sample of L14, whereas $P_{\rm X}$ and $P_{\rm IR}$ were factors of several weaker. The change in the pressure terms between the two \hii\ region populations demonstrates that the dynamical impact of feedback evolves and that direct and dust-processed radiation can be significant at early times (e.g., $\lesssim10^5$ yr; \citealt{arthur06,tremblin14}) in star-forming regions.

Recently, \cite{barnes20} conducted a similar analysis as our work, measuring feedback pressures in \hii\ regions with effective radii $R_{\rm eff} \sim 10^{-3}-10$~pc in the Milky Way Central Molecular Zone. They found that $P_{\rm dir}$ is the dominant feedback term for sources with $R_{\rm eff}<$0.1~pc, whereas $P_{\rm HII}$ takes over the energy budget in larger regions. Although \cite{barnes20} measured $P_{\rm IR}$ for their regions with $R_{\rm eff}\gtrsim$0.5~pc, they did not extend down to smaller radii to estimate $P_{\rm IR}$ because of spatial resolution limitations. However, we note that their estimates of $P_{\rm dir}$ and $P_{\rm HII}$ for sources with $R_{\rm eff}\lesssim$0.5~pc agree within an order-of-magnitude of our results, and their best-fit power-law slopes (of $P_{\rm HII} \propto R_{\rm eff}^{-1.0}$ and $P_{\rm dir} \propto R_{\rm eff}^{-1.5}$) are consistent with our findings (shown in \autoref{fig:pvr}) given the uncertainties. 

\subsection{$P_{\rm IR}$-Dominated Regions}\label{subsec:PIRdom}

From the 81 regions discussed in \autoref{sec:results}, we find that 68 sources have $P_{\rm IR}$ as the dominant pressure term. Three regions are $P_{\rm dir}$-dominated, four regions have comparable $P_{\rm HII}$ and $P_{\rm IR}$, and six regions have comparable $P_{\rm dir}$ and $P_{\rm IR}$.  We report the range of properties for these four groups of sources (with $P_{\rm HII} \sim P_{\rm IR}$, $P_{\rm dir} \sim P_{\rm IR}$, dominant $P_{\rm dir}$, and dominant $P_{\rm IR}$) in \autoref{tbl1}. We note that the $L_{\rm bol}$ of the sources with $P_{\rm HII} \sim P_{\rm IR}$ are in the lower-luminosity end of the sample, whereas the $P_{\rm dir}$-dominated and $P_{\rm dir} \sim P_{\rm IR}$ regions have higher luminosities. The $f_{\rm trap}$ values for the $P_{\rm dir}$-dominated and $P_{\rm dir} \sim P_{\rm IR}$ regions are among the lowest in the sample, while the measurements for the $P_{\rm IR}$-dominated and $P_{\rm HII} \sim P_{\rm IR}$ regions are systematically greater. The range of $R/r_{\rm ch}$ for the regions with $P_{\rm HII} \sim P_{\rm IR}$ is 3.3--11.66, consistent with expectations that regions with $R/r_{\rm ch}>1$ are gas-pressure dominated. As predicted, the vast majority (60 out of 68) of the $P_{\rm IR}$-dominated regions have $R/r_{\rm ch}<1$, though seven have $R/r_{\rm ch}=1-2$ and one has $R/r_{\rm ch}=5.5$. We note that the latter source has a relatively low $f_{\rm trap}=2.9$ and log~$L_{\rm bol}/L_\odot=4.5$, leading to a small $r_{\rm ch}$ of 0.01~pc.

Our results, where 84\% of the young regions are $P_{\rm IR}$-dominated, indicate that radiation pressure can be significant at early times in \hii\ regions. This finding is consistent with the conclusions of \cite{geen20} who employed analytical models of \hii\ regions and showed that $P_{\rm IR}$ and $P_{\rm X}$ are most important when the shells are small (with radii $<$0.1~pc) and at high surface densities. However, our results contrast those from other theoretical studies that evaluated the impact of dust-processed radiation. \cite{rahner17} constructed semi-analytic, one-dimensional models of isolated 10$^{5}$--10$^{8}$~$M_{\sun}$ clouds with radiation (direct and dust-processed), stellar winds, and SNe. They found that $P_{\rm IR}$ is negligible in their low-mass clouds ($\sim$10$^{5}$~$M_{\sun}$) and is only significant in the early phases of the massive GMCs or during recollapse following the initial starburst. \cite{reissl18} noted that photons absorbed and re-emitted by dust deposit little momentum in GMCs because they escape promptly, thus limiting the role of $P_{\rm IR}$ in GMC disruption. However, in agreement with our results in combination with those of L14, they find that radiation pressure decreases as star clusters become more extended and evolved. Our observational results reflect \hii\ regions powered by individual stars or lower-mass clusters at earlier times than considered by \cite{rahner17} and at smaller scales than \cite{reissl18}. Our findings suggest that sub-pc \hii\ regions may have significant indirect radiation pressure, even if it does not lead to GMC disruption.

$P_{\rm IR}$-dominated sources have distinct structure and dynamics (e.g. \citealt{mathews67,mathews69,petrosian72,gail79}) from classical \hii\ regions which are dominated by $P_{\rm HII}$ (e.g., \citealt{stromgren39,savedoff55}). In particular, $P_{\rm IR}$-dominated \hii\ regions are thought to have density gradients in the gas within the regions (e.g. \citealt{dopita03,dopita06,draine11}) and have swept-up shells of gas and dust (e.g. \citealt{draine11,rodriguez-ramirez16}). Dynamically, radiation pressure leads to faster expansion at early times (e.g. \citealt{krumholz09,martinez14,geen20}). Given the differences relative to classical \hii\ regions, our results underscore the importance of including dust and radiative feedback in small-scale massive star formation simulations and high-resolution GMC scale simulations.

\subsection{Application to Sub-Grid Feedback Models in Galaxy- and Star Cluster-Scale Simulations}

As discussed above, one of the primary takeaways from our results is that dust and radiative feedback should not be neglected in small-scale, high-mass star formation and high-resolution, GMC-scale simulations. Quite often, due to the large computational expense of radiative transfer in numerical simulations \citep[e.g., see][]{rosen2017}, the dust-processed radiation pressure is neglected. However, our results indicate that during the early evolution of massive star formation $P_{\rm IR}$ is the dominant feedback mechanism regulating \hii\ region dynamics, at least up to a $R \sim 0.5$ pc, corresponding to the median $r_{\rm ch}$ value found in our sample. 

\begin{figure}
\includegraphics[width=\columnwidth]{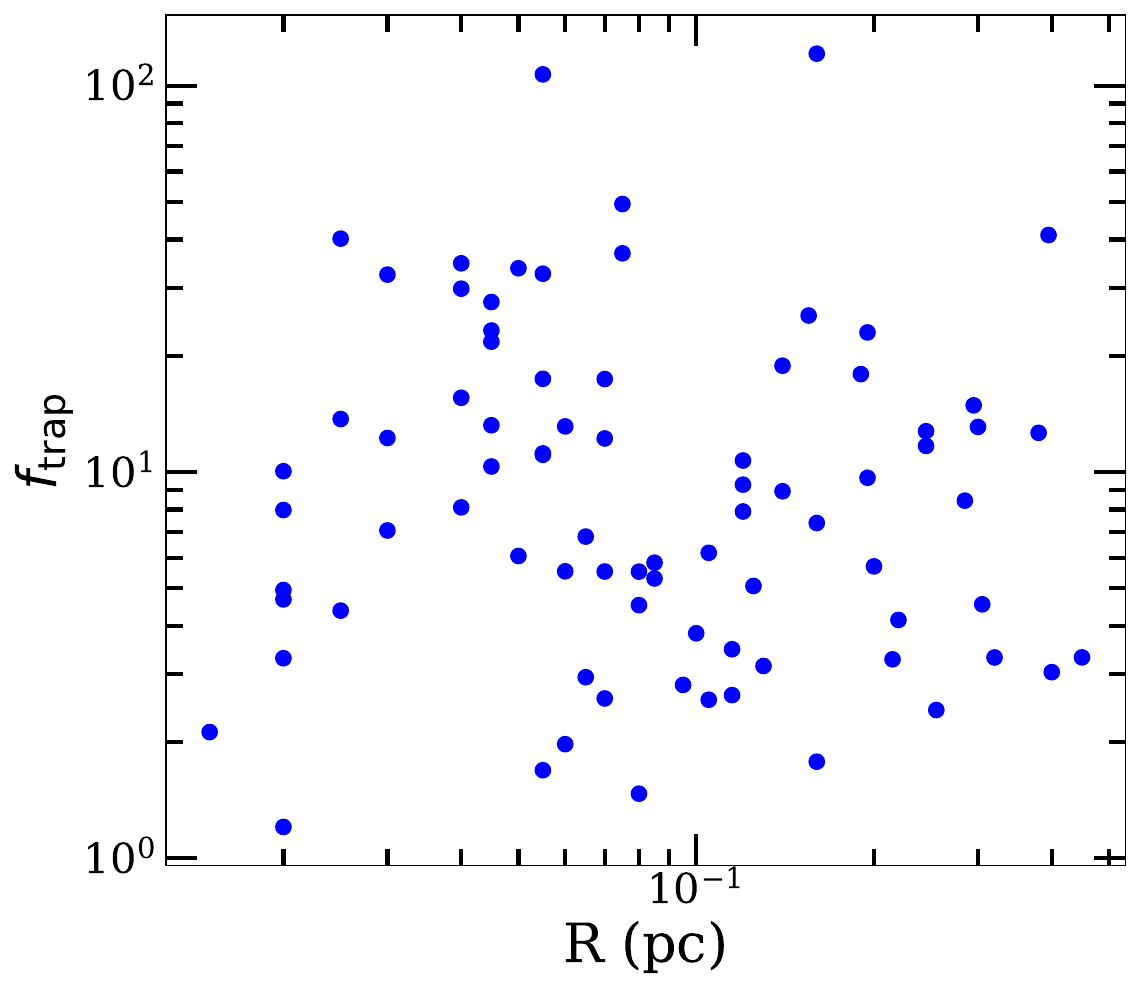}
\caption{Plot of $f_{\rm trap}$ as a function of radius for the young \hii\ region sample. We find no correlation between $f_{\rm trap}$ and $R$, suggesting it is a reasonable approximation to assume a constant $f_{\rm trap}$ at scales of $\lesssim$0.5~pc.}
\label{fig:ftrap_vs_r}
\end{figure}

In the future, simulations can incorporate our observationally-inferred results to model the energy and momentum injection by massive stars into the ISM at these scales by adopting the following sub-grid prescription. Within a radius $R \lesssim 0.5$~pc, the rate of energy ($\dot{E}_{\rm rad}$) and momentum ($\dot{p}_{\rm rad}$) injection in nearby gas and dust by massive stars is

\begin{eqnarray}
    \dot{E}_{\rm rad} &=& f_{\rm trap} L_{\rm bol} \\
    \dot{p}_{\rm rad} &=& \frac{\dot{E}_{\rm rad}}{c}
\end{eqnarray}

\noindent
where $L_{\rm bol} = L_{\rm \star }+ L_{\rm acc}$ is the sum of the stellar luminosity and accretion luminosity. Here, $f_{\rm trap}$ takes into account both the direct radiation and indirect radiation pressure enhancement due to reprocessing by dust. We suggest using a value of $f_{\rm trap} \sim 8$ following the median value obtained in our sample. Given that we do not find a correlation between \hii\ region size and $f_{\rm trap}$ (as shown in \autoref{fig:ftrap_vs_r}), we conclude that using a constant value for $f_{\rm trap}$ is a reasonable approximation. However, we note that this $f_{\rm trap}$ does likely depend on other physical parameters, e.g. the dust-to-gas ratio, metallicity, and the optical depth per unit dust mass.

\subsection{Comparison to Gravitational Pressure}

Since these regions are compact they may be actively accreting material and gravitationally collapsing.  We test this hypothesis by calculating the gravitational pressure ($P_{\rm grav}$) to compare to our feedback pressures as calculated above. From our models discussed in \autoref{sec:methods}, we estimate the gravitational pressure as

\begin{equation}
    P_{\rm grav} = \frac{G M^2}{R V}
\end{equation}

\noindent 
where $G$ is the gravitational constant, $M$ is the total mass within radius $R$ (including the star and gas), and $V = 4 \pi R^3/3$ is the volume of the \hii\ region.

\begin{figure}
\includegraphics[width=\columnwidth]{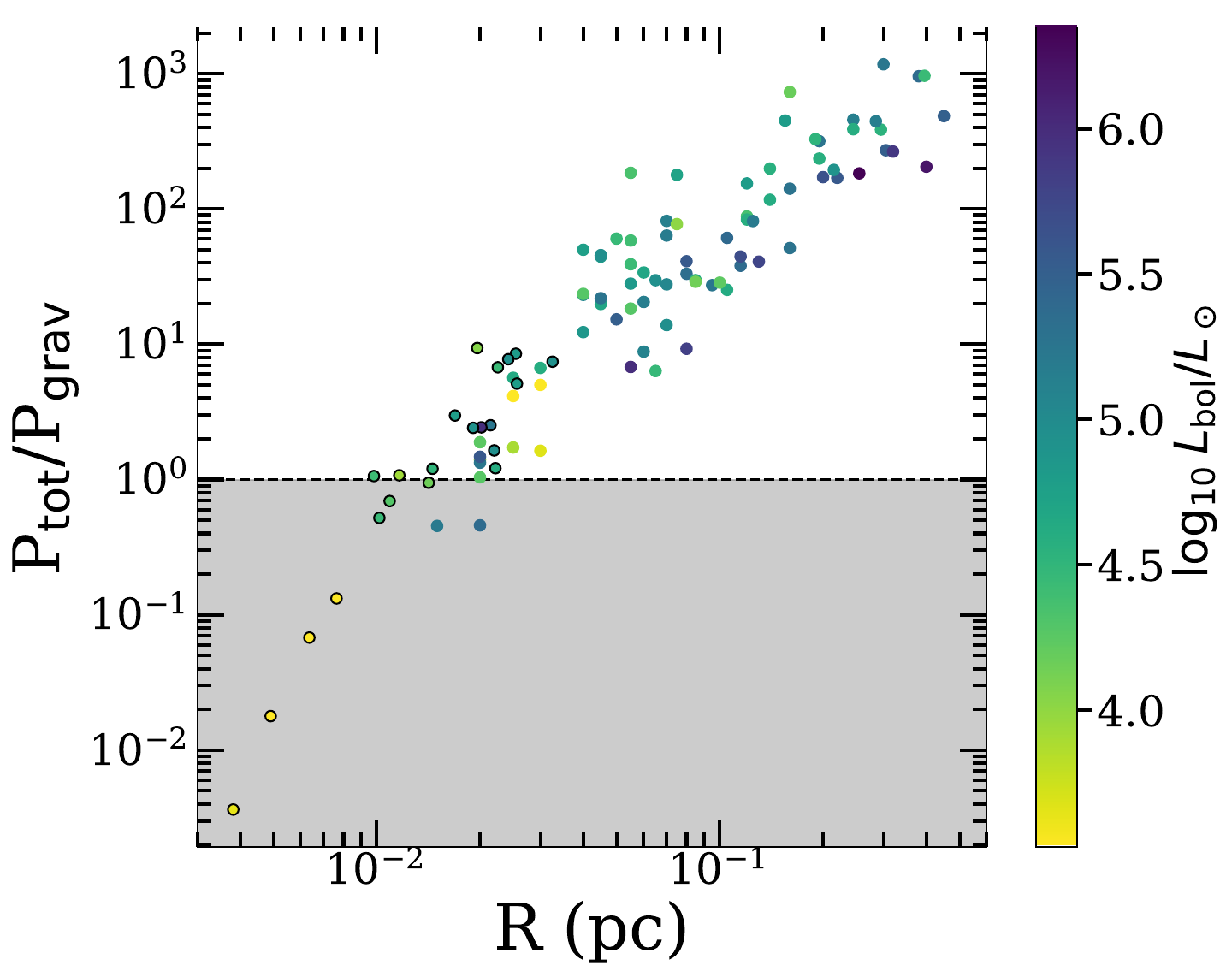}
\caption{Plot of $P_{\rm tot}/P_{\rm grav}$ versus radius $R$, where $P_{\rm tot} = P_{\rm HII} + P_{\rm dir} + P_{\rm IR}$.  The circles are filled based on their $L_{\rm bol}$ as shown in the colorbar. Symbols with a black outline correspond to those with upper-limits on $R$. The shaded area represents the parameter space where $P_{\rm grav}>P_{\rm tot}$, and thus the sources there are still collapsing due to gravity. Most of these $P_{\rm grav}>P_{\rm tot}$ sources have low $L_{\rm bol}$, which indicates smaller or younger stars. The vast majority of our young \hii\ region sample have $P_{\rm tot} > P_{\rm grav}$, suggesting the sources are expelling gas due to feedback.}
\label{fig:pgrav}
\end{figure}

We compare $P_{\rm grav}$ to the total pressure from feedback, $P_{\rm tot} = P_{\rm HII} + P_{\rm dir} + P_{\rm IR}$, in \autoref{fig:pgrav} by plotting their ratio as a function of $R$.  The regions with $P_{\rm tot} > P_{\rm grav}$ represent those dominated by feedback, while those with $P_{\rm tot} < P_{\rm grav}$ (in the shaded region of \autoref{fig:pgrav}) are collapsing due to gravity. The figure shows that for all but the most compact of our \hii\ regions, feedback dominates over gravity, and thus is capable of ejecting gas.  For the few regions that are collapsing due to gravity, we interpret this result as the late stages of accretion onto the star as these regions are the smallest in our sample and have some of the lowest $L_{\rm bol}$ measurements, which could indicate the stars are still growing. We expect that the feedback pressures will increase over time and then drive \hii\ region expansion. The existence of accretion-confined \hii\ regions such as those we have found has long been predicted (e.g., \citealt{keto02,keto03}).

\section{Conclusions} \label{sec:conclusions}

In this work, we used radio, IR, and X-ray data to assess the dynamical impact of stellar feedback mechanisms in a sample of 106 resolved, young \hii\ regions with radii $<$0.5~pc.  We measured the pressures associated with direct radiation ($P_{\rm dir}$), dust-processed radiation ($P_{\rm IR}$), photoionization heating ($P_{\rm HII}$), and stellar winds ($P_{\rm X}$). We found that $P_{\rm IR}$ is statistically significantly dominant for 84\% (68 out of 81) of the regions, and by comparison, the median $P_{\rm dir}$ ($P_{\rm HII}$) is 17\% (11\%) of the median $P_{\rm IR}$ in our sample.  We set upper limits on $P_{\rm X}$ due to the lack of X-ray detections, and these limits are comparable to the measured $P_{\rm IR}$ values.

Our young \hii\ region results contrast with those of L14, who analyzed evolved sources in the LMC and SMC that were mostly $P_{\rm HII}$-dominated. Our sample yielded radial pressure dependences of $P_{\rm HII} \propto R^{-0.74\pm0.07}$, $P_{\rm dir} \propto R^{-1.36\pm0.16}$, and $P_{\rm IR} \propto R^{-1.43\pm0.15}$. Using these relations, the transition radius from $P_{\rm IR}$-dominated and $P_{\rm HII}$-dominated regions would be at $\sim$3 pc. We found a median $f_{\rm trap}$ of $\sim$8 without any radial dependence for regions $\lesssim$0.5~pc in size, suggesting this value can be adopted in sub-grid feedback models in galaxy- and star cluster-scale simulations. We compared the total pressure $P_{\rm tot}$ to the gravitational pressure $P_{\rm grav}$ in our sources, and we showed that only the smallest \hii\ regions are dominated by $P_{\rm grav}$. This result indicates that for the majority of our sources, the feedback pressure is sufficient to expel gas from the regions. 

In the future, observations of \hii\ regions with radii of $\sim$0.5--3~pc can fill the gap between the young sources considered here and the evolved sample analyzed by L14. That work will enable stronger constraints on the scale where \hii\ regions transition from $P_{\rm rad}$-dominated to $P_{\rm HII}$-dominated.  

\acknowledgements

G.M.O. would like to thank James W. Johnson, Adam Leroy, Todd Thompson, and the OSU Galaxy/ISM Group Meeting for useful discussions during the preparation of this work. G.M.O. and L.A.L. are supported by a Cottrell Scholar Award from the Research Corporation of Science Advancement. A.L.R. acknowledges support from NASA through Einstein Postdoctoral Fellowship grant number PF7-180166 awarded by the Chandra X-ray Center, which is operated by the Smithsonian Astrophysical Observatory for NASA under contract NAS8-03060. M.R.K. acknowledges support from the Australian Research Council through its Future Fellowship and Discovery Projects schemes (awards FT180100375 and DP190101258. This work is based in part on observations made with the \textit{Spitzer Space Telescope}, which is operated by the Jet Propulsion Laboratory, California Institute of Technology under a contract with NASA. This publication makes use of data products from the Two Micron All Sky Survey, which is a joint project of the University of Massachusetts and the Infrared Processing and Analysis Center/California Institute of Technology, funded by the National Aeronautics and Space Administration and the National Science Foundation. This paper made use of information from the Red MSX Source survey database at http://rms.leeds.ac.uk/cgi-bin/public/RMS\_DATABASE.cgi which was constructed with support from the Science and Technology Facilities Council of the UK. L.A.L. and A.L.R. would also like to thank their “coworkers,” Slinky, Kerfuffle, and Nova for ``insightful conversations" and unwavering support at home while this paper was being written during the coronavirus pandemic of
2020. Their contributions were not sufficient to warrant co-authorship due to excessive napping.\footnote{Because they are cats.} 

\noindent
{\it Facilities}: {\it Chandra}, {\it Spitzer}

\software{CIAO (v4.7; \citealt{fru06}), HYPERION (v0.9.10; \citealt{robitaille11}), SEDfitter (\citealt{robitaille07}),
}

\nocite{*}
\bibliographystyle{aasjournal}
\bibliography{feedback}

\end{document}